\documentclass[manuscript]{aastex}
\usepackage{amssymb}
\usepackage{graphicx}

\def\half{{\textstyle{1\over2}}}

\def\quarter{{\textstyle{1\over4}}}
\newcommand\pa[2]{\frac{\partial{#1}}{\partial{#2}}}
\def\Rs{R_{\odot}}
\def\Ru{R_{\textrm{\tiny top}}}
\begin{document}
\title{Solar flows and their effect on frequencies of acoustic modes} 

\author{Piyali Chatterjee\thanks{Now at NORDITA, Stockholm 106 91, Sweden}}
\email{piyalic@tifr.res.in}
\and 
\author{H. M. Antia}
\email{antia@tifr.res.in}
\affil{Department of Astronomy and Astrophysics, Tata Institute of Fundamental
Research, Homi Bhabha Road, Mumbai 400005, India}
\date{}

\begin{abstract}
We have calculated the effects of large scale solar flows like the meridional
circulation, giant convection cells and solar rotation on the helioseismic
splitting coefficients using quasi-degenerate perturbation theory (QDPT).
Our investigation reveals that the effect of poloidal flows like the large scale
meridional circulation are difficult to detect in observational data of the
global acoustic modes since the frequency shifts are much less than the errors.
However, signatures of large scale convective flows may be detected if their
amplitude is sufficiently large by looking for frequency shifts due to nearly
degenerate modes coupled by convection.
In this comprehensive study, we attempt to put limits on the magnitude of flow
velocities in giant cells by comparing the splitting coefficients obtained from
the QDPT treatment with observational data.
\end{abstract}
\maketitle

\section{Introduction}
Solar convection is believed to be organized in a variety of spatial and
temporal scales ranging from granules (size $\sim 1$ Mm, 0.2 hr lifetime),
mesogranules (size $\sim 10$ Mm, 3 hr lifetime), supergranules (size $\sim 30$
Mm, 1 day lifetime) to the  giant cells (size $\ge 100$ Mm, 1 month lifetime).
The power spectra of convective velocities show distinct peaks representing
granules and supergranules but no distinct features at wavenumbers
representative of mesogranules or giant cells (Wang 1989; Chou et al.~1991;
Straus, Deubner, \& Fleck 1992; Straus \& Bonaccini 1997; Hathaway et al.~2000).
Numerical simulations of solar convection routinely show the existence of
mesogranules and giant cells (Miesch et al.~2000; Miesch et al.~2008).
In this study we will only concentrate on the theoretical frequency shifts in
global surface gravity and acoustic modes, denoted f- and p-modes respectively
due to interaction with large scale flows namely the giant cells, the
meridional circulation and rotation.
The meridional circulation, believed to play an important role in magnetic flux
transport by Solar dynamo modelers (Choudhuri, Sch\"ussler \& Dikpati 1995;
Dikpati \& Charbonneau 1999; Chatterjee, Nandy \& Choudhuri 2004) was observed
at the solar surface by Doppler measurements of photospheric lines (Duvall 1979;
LaBonte \& Howard 1982; Komm, Howard \& Harvey 1993; Hathaway 1996) and later by
using techniques of ring diagram and time-distance helioseismology (Giles et
al.~1997; Basu, Antia \& Tripathy 1999; Gonz\'alez Hern\'andez et al.~1999).
These studies can only measure the flow velocity in the near surface layers.
This flow having a maximum surface velocity of 30 m s$^{-1}$ is apparently
poleward all the way from the equator to the poles at the Solar surface.
Conservation of mass requires existence of a return flow advecting mass from the
poles to the equator somewhere below the surface.
The dynamo models assume that the return flow occurs near the base of the
convection zone, but there is no direct evidence about where the return flow
is located.
It is believed that the only hope of detecting such a return flow is by
analyzing the effect of the meridional circulation on the global helioseismic
acoustic modes (Roth \& Stix 2008). Their investigation using quasi degenerate
perturbation theory showed that theoretical p-mode frequency shifts due to
meridional circulation would be $\sim 0.1~\mu$Hz.
Such large frequency shifts may not be expected from the observed magnitude of
meridional flows.
For the sake of comparison, the equatorial rotational velocity at the Solar
surface is $\sim 2000$ m s$^{-1}$ and gives rise to frequency shifts $\sim
450\times m$ nHz as a linear function of azimuthal order $m$.
Further, since the first order contribution from meridional flow calculated
using the degenerate perturbation theory vanishes, one would expect the effect
to be even smaller than what is suggested by the velocity. Thus it is necessary
to reexamine these calculations.

Quasi degenerate perturbation theory (QDPT) was applied to calculate shifts due
to flows and asphericity in Solar acoustic frequencies by Lavely \& Ritzwoller
(1992). We have used the same formulation to calculate the p-mode frequency
shifts due to differential rotation, meridional circulation and giant cells of
convection.
In contrast to results of Roth \& Stix (2008), our results indicate a maximum
shift of about 1 nHz for a similar meridional circulation encompassing the
entire solar convection zone. Traditionally, only degenerate perturbation theory
(DPT) has been used to calculate the effect of rotation on the p-modes.
Woodard (1989) and Vorontsov (2007) applied QDPT to calculate mixing of
eigenfunctions due to rotational coupling of p-modes and concluded that the
expected line profiles of the composite modes in the Solar power spectra are not
distorted significantly. In this work, we use a different approach to examine if
the use of quasi degenerate perturbation
theory introduces significant corrections in the frequency shifts over that
obtained from degenerate
perturbation theory. It is necessary to check this as inversion techniques
for calculating rotation rate in the solar interior are based on degenerate
perturbation theory. Helioseismic inversion for rotation is also insensitive to
the N--S asymmetric component of rotation.
This is because in the degenerate perturbation theory, this effect also gives
zero contribution.
The N--S antisymmetric component of rotation has been studied near the surface
and is found to be small. In this work we wish to examine its effect to check
if its contribution to frequency shift are indeed small.

The giant convective cells have always been elusive to observations at the
solar surface.
In the past there have been studies which have failed to detect giant cell
motions (LaBonte, Howard \& Gilman 1981; Snodgrass \& Howard 1984; Chiang, Petro
\& Foukal 1987) as well as those hinting at their existence (Cram, Durney \&
Guenther 1983; Hathaway et al.~1996; Simon \& Strous 1997).
With the availability of Dopplergrams from SOHO/MDI, it became possible to study
such long-lived and large scale features more reliably.
Beck, Duvall \& Scherrer (1998) were able to detect giant cells at the solar
surface with large aspect ratio ($\sim 4$) and velocities $\sim 10$ m s$^{-1}$
using the MDI data. Hathaway et al.~(2000) used spherical harmonic spectra from
full disk measurements to detect long-lived power at $l \le 64$. This means that
the spatial extent of the features were $\sim 2\pi\Rs/\sqrt{l(l+1)} \ge 70$~Mm.
Later Ulrich (2001) identified these long-lived features not as giant cells but
higher order components of the torsional oscillation. Interestingly it was
earlier proposed by Snodgrass \& Wilson (1987) that torsional oscillations occur
due to modulation of differential rotation by the Coriolis force arising from
meridional motions of giant cells. 

The elusiveness of the giant cell features in the surface velocities lead
modelers to speculate about the subsurface nature of the giant cell motions
(Latour, Toomre \& Zahn 1981). Roth \& Stix (1999, 2003) used QDPT to calculate
the effect of giant cells on p-modes and claimed that giant cells could be found
by modeling the asymmetries and line broadening in the Solar power spectrum.
They claimed that finite line width of the multiplets would limit the detection
of the frequency splittings to vertical velocity amplitude of 100 m s$^{-1}$ or
larger. However, Roth, Howe \& Komm (2002) claimed that giant cells could be
detected with current inversion methods of global helioseismology as long as
they exceed an amplitude of 10 m s$^{-1}$.
Usual inversion procedure for rotation neglects giant cells and assumes
that the odd splitting coefficients arise only from rotation. If there
are additional contributions to these coefficients, rotation inversions
would not give correct results.
According to Roth, Howe \& Komm (2002) the effect of giant cells would appear as
distortions in the rotation inversion if their velocities are $\ge 10$ m
s$^{-1}$.
However, using the same technique of QDPT we find that with the present global
helioseismic data we can put an upper limit of 50--100 m s$^{-1}$ on the
vertical velocity associated with giant cell flows.
This is mainly because our calculations show much smaller effects from
giant cells.

This paper is organized as follows. In \S 2 we define zonal and poloidal flows
inside the Sun, while \S3 discusses the quasi degenerate perturbation theory,
its differences from degenerate theory as well as the details of our
calculations. We present our theoretical results and compare them with present
f- and p-mode data from both GONG and SOHO/MDI instruments in \S4. Finally
conclusions are described in \S5.

\section{Zonal and Poloidal Solar Flows}
Following earlier works (Lavely \& Ritzwoller 1992; Roth, Howe \& Komm 2002) we
express the velocity field in terms of spherical harmonics. For completeness, we
give the expression here again.
\begin{eqnarray}
\label{eq:flow}
\nonumber
{\bf v}(r,\theta,\phi) &=& \mathrm{Re}[u_s^t(r) Y_s^t(\theta, \phi)
{\bf{\hat{r}}} + v_s^t(r) {\bf \nabla}_{h}Y_s^t(\theta, \phi) - \\
&&\qquad w_s^t(r) {\bf{\hat{r}}}\times {\bf\nabla}_{h}Y_s^t(\theta, \phi)] .
\end{eqnarray}
The quantities $u_s^t, v_s^t$ and $w_s^t$ determine the radial profiles of the
flows and ${\bf\nabla}_{h}$ is the horizontal gradient operator. The
$\mathrm{Re}$ refers to using only the real part of the spherical harmonics as
in,
\begin{equation}
\mathrm{Re}[Y_s^t(\theta,\phi)]=\cases{[Y_s^{-t}(\theta,\phi)+Y_s^{t}(\theta,
\phi)]/2 & if $t$ is even, \cr 
\noalign{\smallskip}
[Y_s^{-t}(\theta,\phi)-Y_s^{t}(\theta,\phi)]/2 & if $t$ is odd.\cr}
\end{equation}
The first two terms in equation~(\ref{eq:flow}) define the poloidal component of
the flow whereas the last term is the toroidal component.
By the poloidal component, we imply the meridional and non-zonal toroidal flows
(average over $\phi$ direction is zero) e.g., (i) the meridional circulation
which carries mass poleward near the surface and sinks near the poles and (ii)
the giant convection cells, respectively.
These flows are also called large scale flows to distinguish them from other
small scale flows like the turbulent eddies which are of the size smaller than
the typical scale of global modes used in helioseismology.
In presence of only the poloidal flow ($w_s^t = 0$) we can apply the equation of
mass conservation
$\nabla.(\rho_{0}{\bf v}) = 0$ to get a relation between $u_s^t(r)$ and
$v_s^t(r)$ e.g.,
\begin{equation}
\label{eq:vst}
v_s^t(r) = \frac{1}{r}\pa{}{r} \left[\frac{\rho_{0} r^2
u_s^t(r)}{s(s+1)}\right]\;.
\end{equation}
Here $\rho_0(r)$ is the density in a spherically symmetric solar model.
So now it only remains to choose $u_s^t(r)$ appropriately and $v_s^t(r)$ will be
determined by equation~(\ref{eq:vst}). We choose the radial profile of
$u_s^t(r)$ as given by equation~(19) of Roth, Howe \& Komm (2002): 

\begin{equation}
\label{eq:profile}
u_s^t(r)=\cases{u_0\frac{4(\Ru-r)(r-r_b)}{(\Ru-r_b)^2} & if $r_b\le r\le \Ru$,
\cr 0 & otherwise.\cr}
\end{equation}
Here $r_b$ and $\Ru$ define the boundaries of region where the flow is
confined.
The tangential component $v_s^t$ also vanishes outside these boundaries.
Further, because of very small density scale height near the solar surface,
$v_s^t$
increases very rapidly near the top boundary and its maximum value depends
on the choice of $\Ru$. For the meridional flow such a behavior has
not been seen in the actual observed profile (e.g., Basu, Antia \& Tripathy
1999).
In most of our computations we use $r_b = 0.7\Rs$, the approximate
position of the base of the convection zone, while $\Ru$ is taken to
be the top of the solar model used in this work, which is around
$1.0017R_\odot$. With this choice most of the steep variation in $v_s^t$ occurs
above $r=\Rs$ and the variation is probably more reasonable below
the solar surface.

The radial profile of the vertical velocity and the horizontal velocity for the
upper boundary at $\Ru=\Rs$ is shown in Figure~\ref{fig:profile}.
We wish to point out that Roth, Howe \& Komm (2002) have used an upper boundary
at $\Ru=0.99\Rs$ in equation~(\ref{eq:profile}).
This avoids the region with small density scale height and hence the maximum
value of $v_s^t(r)$ is smaller.
In this work we have used a standard solar model with the OPAL equation of
state (Rogers \& Nayfonov 2002) and OPAL opacities (Iglesias \& Rogers 1996) and
use the formulation due
to Canuto \& Mazzitelli (1991) to calculate the convective flux.
The flow obtained using equation~(\ref{eq:profile}) consists of only one cell in
the radial direction.
Following Roth, Howe \& Komm (2002) we also use giant cells with a horizontal
wave number $s=8$.
The streamlines of the flow for $s=8, t=8$ as well as for $s=8,t=0$ are given in
Figure~7 of Roth, Howe \& Komm (2002).

The differential rotation constitutes the zonal toroidal flows with $t=0$ in
last term of equation~(\ref{eq:flow}). The radial and angular dependence of
uniform rotation is given by $w_1^0(r)\partial_{\theta}Y_{1}^0(\theta,\phi)$ and
that of differential rotation with $s=3$ is given by
$w_3^0(r)\partial_{\theta}Y_{3}^0(\theta,\phi)$ respectively.
Rotation affects the p-modes in two ways: (i) Coriolis force which is linear
in rotation rate, $\Omega$ and (ii) centrifugal force which is quadratic in
$\Omega$.
To first order, the linear term gives only odd splitting coefficients, while
the centrifugal term gives only even splitting coefficients.
The effect of centrifugal force is of second order compared to (i) and has been
calculated by Antia et al.~(2000).
In this work we would consider only the linear term due to Coriolis force,
or advection in an inertial frame, though we apply QDPT to calculate
second order corrections due to rotation.

\section{Perturbation of f- and p-modes}
The eigenfunctions and eigenfrequencies for a standard solar model can be
calculated using perturbative or variational methods. We follow the approach of
Lavely \& Ritzwoller (1992) and use the quasi-degenerate perturbation theory, as
opposed to degenerate perturbation theory, to determine these quantities.
For meridional flow this is necessary as the degenerate perturbation gives
zero frequency shifts.
Let the eigenfunctions of the operator $\mathcal{L}_0$ governing the seismic
oscillations in a standard solar model be called the non-rotating standard solar
model (NRSSM) eigenfunctions. We shall see in \S3.1 how to appropriately choose
the set of NRSSM eigenfunctions for quasi-degenerate and degenerate treatments.

\subsection{Quasi-degenerate perturbation theory}
 The difference between QDPT and DPT lies in the choice of the set of NRSSM
eigenfunctions  (${\bf s_k}$), whose linear combinations are used to express the
perturbed eigenfunctions 
 (${\bf s_k'}$) e.g.,
\begin{equation}
\label{eq:lin}
{\bf s_k'} = \sum_{k'\in K} a_{k'} {\bf s_{k'}}.
\end{equation}
In DPT only the eigenfunctions from the standard solar model which are {\em
exactly} degenerate constitute the eigenspace $K$. In contrast to this in QDPT
the set consists of all eigenfunctions of the standard solar model with {\em
nearly} degenerate frequencies.
It is well known that simple harmonic oscillators couple strongly only if the
natural uncoupled frequencies of the oscillators are nearly degenerate. Hence it
is important to determine which of the NRSSM eigenfunctions should be included
in the analysis.
Let the eigenspace of the suitable eigenfunctions be called $K$. Qualitatively,
the quasi-degenerate condition implies that we include an eigenfunction ${\bf
s}_{nlm}(=\xi_{nl}(r)Y_{l}^m{\bf{\hat{r}}}+\eta_{nl}(r){\bf{\nabla}}_{h}Y_l^m)
\in K$ iff $|\omega_{nlm} -\omega_{\textrm{ref}}| < \epsilon$, where $\epsilon$
is the radius of the neighborhood about the central frequency.
We have generally used $\epsilon=100\;\mu$Hz in this work. This is comparable
to asymptotic spacing between two consecutive modes in radial order $n$.
Moreover no significant change in calculated frequency shifts is observed for
$\epsilon > 100\; \mu$Hz.
Here $\omega_{\textrm{ref}}$ is the frequency of the central eigenmode for which
we require the frequency shift. The size of the eigenspace $K$ will be
determined by the (i) desired level of accuracy which is set by $\epsilon$ and
(ii) the selection rules of the perturbation operator.

In the NRSSM the frequencies are independent of the azimuthal order $m$ and
this degeneracy is lifted by departures from spherical symmetry due to rotation
or magnetic field. Helioseismic data traditionally give the splitting
coefficients for all modes that are detected. These coefficients are defined
by (e.g., Ritzwoller \& Lavely 1991)
\begin{equation}
\label{eq:m}
\omega_{nlm} = \omega_{nl} + \sum_{q}a_q^{(nl)}\mathcal{P}^{l}_q(m),
\end{equation}
where $\omega_{nl}$ is the mean frequency of the multiplet,
$\mathcal{P}_q^{l}(m)$ are the orthogonal polynomials of degree $q$ and
$a_q^{(nl)}$'s are the so called splitting coefficients.

The equations of motion for a mode $k$ with eigenfrequency $\omega_k$ for a
NRSSM and a model perturbed by addition of differential rotation and/or large
scale flow can be respectively represented by,
\begin{eqnarray}
\label{eq:qdpt1}
{\mathcal{L}_0}{\bf s_k}&=&-\rho_0\omega_k^2 {\bf s_k},\\
\label{eq:qdpt2}
{\mathcal{L}_0}{\bf s'_k}+{\mathcal{L}_1}{\bf s'_k}&=&-\rho_0{\omega'_k}^2 {\bf
s'_k},
\end{eqnarray}
where $\omega_k'$ is the perturbed frequency and ${\bf s'_k}$ given by
equation~(\ref{eq:lin}) is the perturbed eigenfunction.
Taking scaler product with ${\bf s_j}$ in equation~(\ref{eq:qdpt2}) and using
the notation ${\mathcal H}_{jk'} = - \int {\bf s_{j}}^{\dagger} {\mathcal L}_1
{\bf s_{k'}} dV$ and the definition  $\mathcal{L}_1 {\bf s_k} =
-2i\omega_{\textrm{\scriptsize ref}}\rho_0({\bf v.\nabla)s_k}$, we obtain the
matrix eigenvalue equation,
\begin{equation}
\label{eq:qdpt3}
 \sum_{k' \in K} \left\lbrace {\mathcal H}_{jk'} + \delta_{k'j} (\omega_{k'}^2 -
\omega_{\textrm{ref}}^2)\right\rbrace a_{k'} = 
({\omega'_k}^2-\omega_{\textrm{ref}}^2)a_{j},
\end{equation}
with, eigenvalue $\lambda = ({\omega'_k}^2- \omega_{\textrm{ref}}^2)$ and
eigenvector $X_{j} = \left\lbrace a_{j} \right\rbrace$.
Here $\omega_\textrm{ref}$ is a reference frequency which approximates
$\omega_k'$. In this work we use $\omega_\textrm{ref}=\omega_k$, the frequency
of the mode being perturbed.
It will be clear from \S3.2 that the perturbation matrix
$[\mathcal{H}_{n'n,l'l}^{m'm}]$ is Hermitian for both differential rotation and
poloidal flows.
For example, if we consider only two modes with frequencies $\omega_1$ and
$\omega_2$,
then quasi degenerate perturbation theory gives the following coupling matrix
for the mode $\omega_2$,
\begin{equation}
\left[\begin{array}{cc}
{\mathcal H}_{11} -\Delta & {\mathcal H}_{12}\\
{\mathcal H}_{21} & {\mathcal H}_{22}\\
\end{array} \right]
\end{equation}
with $\Delta= \omega_2^2 - \omega_1^2$. We are interested in the eigenvalues
$\lambda$ of $[\mathcal{H}_{ij}]$ corresponding to perturbation in $\omega_2$
which may be easily shown to be,
\begin{equation}
\lambda  \sim  {\mathcal H}_{22} -\frac{{|\mathcal H}_{12}|^2}{{\mathcal H}_{11}
-{\mathcal H}_{22} -\Delta}\;.
\label{eq:qdp4}
\end{equation}
Let $\omega_2'$ be the modified frequency of the mode which initially has a
frequency $\omega_2$. Then from equation~(\ref{eq:qdp4}) we have,
\begin{equation}
\omega'_2 \sim \left[\omega_2^2+{\mathcal H}_{22} -\frac{|{\mathcal
H}_{12}|^2}{{\mathcal H}_{11} -{\mathcal H}_{22} -\Delta}\right]^{\half}\;.
\end{equation}
If DPT were used then the last term involving off-diagonal term
$\mathcal{H}_{12}$
will not be present and hence this term gives the correction arising from
using the QDPT. It is clear that this correction can be large if the two
modes are nearly degenerate.

If $|H_{12}| \approx |\Delta|$, then the frequency shift due to use of QDPT
would be
of order of $\Delta/\omega_2$, which is comparable to the difference
$\omega_2-\omega_1$. On the other hand if $|H_{12}|\ll|\Delta|$, then the
frequency shift would be much less.
For a typical p-mode the spacing
$\omega_2-\omega_1$ is of order of a few $\mu$Hz.
For a rotation velocity of 2000 m s$^{-1}$, the frequency shift is of the
order of $450m$ nHz, which would define the
magnitude of diagonal terms. Thus for meridional flow with averaged velocity of 20 m
s$^{-1}$
we may expect the matrix elements $H_{12}/\omega_2\sim 4.5l$ nHz.
However, in this case the diagonal elements are zero and we need the
off-diagonal element which involves cross
product of two different eigenfunctions in the integration (cf.,
Eq.~\ref{eq:matrot2})
and (for the same velocity) this integral would be more than order of magnitude
smaller than that in
diagonal term for rotation. Thus the magnitude of $H_{12}$ would be an
order of magnitude less, giving a frequency shift of order of
$0.2l^2/(\omega_2-\omega_1)$ nHz, taking the maximum value of $m$.
For $l\approx100$ this would give
a shift of order of a few nHz. This is much smaller than that obtained by
Roth \& Stix (2008).

In the next section we shall give the expression for the perturbation matrix
elements $[\mathcal{H}^{mm'}_{nn',ll'}]$ due to rotation as well as poloidal
flows to be used for solving the eigenvalue equation~(\ref{eq:qdpt3}).
\subsection{Calculating the perturbation matrix}
The Wigner-Eckart theorem (equation~5.4.1 of Edmonds 1960) states that the
general matrix element of any tensor perturbation operator can be expanded in
terms of Wigner 3$j$ symbols whose coefficients of expansion are independent of
azimuthal order $m$ and $m'$.
\begin{equation}
\label{eq:genmat}
\mathcal{H}^{m'm}_{n'n, l'l} = (-1)^{m'} \left(\begin{array}{ccc}
l' & s & l \\
-m' & t & m
\end{array}\right) (n'l'||\mathfrak{L}_s^{t}||nl).
\end{equation}

The coefficient of the Wigner 3$j$ symbol in equation~(\ref{eq:genmat}) due to
coupling by the differential rotation is given by,
\begin{eqnarray}
\label{eq:matrot}
\nonumber
(n'l'||\mathfrak{L}_s^{t}||nl) =  8 \pi
\omega_{\textrm{ref}}(1-(-1)^{l+l'+s})\times \\
\gamma_{l}\gamma_{l'}\int_{0}^{\Rs} \gamma_s w_s^0 \rho r^2 dr T_s(r),
\end{eqnarray}
where,
\begin{eqnarray}
\label{eq:matrot2}
\nonumber
T_s(r) &=& -\frac{1}{r}\left\lbrace \xi'\eta  +\eta'\xi-\xi'\xi
-\phantom{\frac{1}{2}}\right.\\ \nonumber
&&\left. \frac{1}{2}\eta'\eta[l(l+1)+l'(l'+1)-s(s+1)]\right\rbrace \times \\
&&\qquad\Gamma_{l}\Gamma_{l'}\left(\begin{array}{ccc}
l'& s & l \\ 
-1&  0 & 1
\end{array}\right),
\end{eqnarray}
$\Gamma_q = \sqrt{q(q+1)/2}$ and $\gamma_q = \sqrt{(2q+1)/4\pi}$. $\xi$ and
$\eta$ are radial and horizontal components of the eigenfunction defined in
\S3.1. 

Similarly we have derived the coefficient of the Wigner 3$j$ symbol in
equation~(\ref{eq:genmat}) due to coupling by the large scale poloidal flow,
\begin{eqnarray}
\label{eq:matele1}
\nonumber
(n'l'||\mathfrak{L}_s^{t}||nl) =  8 \pi i \omega_{\textrm{ref}}
(1+(-1)^{l+l'+s}) \times \\
\gamma_{l}\gamma_{l'}\int_{0}^{\Rs} \gamma_s u_s \rho r^2 dr \left[ R_s -
\pa{H_s}{r}\right],
\end{eqnarray}
where,
\begin{eqnarray}
\label{eq:matele2}
\nonumber
R_s(r) &=& \quarter \left(\xi'\pa{\xi}{r}-\pa{\xi'}{r} \xi \right)
\left(\begin{array}{ccc}
l' & s & l\\ 
0 & 0 & 0
\end{array}\right) +\\ 
&& \qquad \half \Gamma_{l}\Gamma_{l'}\left(\eta'\pa{\eta}{r}-\pa{\eta'}{r} \eta
\right)\left(\begin{array}{ccc}
l' & s & l\\
-1 & 0& 1\end{array}\right), 
\end{eqnarray}
\begin{eqnarray}
\label{eq:matele3}
\nonumber
H_s(r)&=&\half [l(l+1)-l'(l'+1)]\times \\
&&\qquad\left[\half \xi \xi'\left(\begin{array}{ccc}
l' & s& l \\
0 & 0 & 0\end{array}\right) - \eta \eta'
\Gamma_{l}\Gamma_{l'}\left(\begin{array}{ccc}
l' & s & l \\
-1 & 0 & 1\end{array}\right)\right] \\ \nonumber
&&\qquad-\eta'\xi\Gamma_{l'}\Gamma_{s}\left(\begin{array}{ccc}
l' & s & l \\
-1 & 1 & 0\end{array}\right)+\xi'\eta
\Gamma_{l}\Gamma_{s}\left(\begin{array}{ccc}
l' & s&l \\
0 & 1 &-1\end{array}\right).
\end{eqnarray}
The selection rules for the Wigner 3$j$ symbol to be non zero are,
\begin{enumerate}
\item $ p/2 \ge {\textrm{max}}(l, l', s)$, where  $p=l+l'+s$,
\item $m-m'+t =0$.
\end{enumerate}

The selection rule for non trivial reduced matrix elements due to differential
rotation is given by,
\begin{enumerate}
\setcounter{enumi}{2}
\item $l+l'+s$ must be odd.
\end{enumerate}
This means that the $\partial_{\theta}Y_3^0$ component of differential rotation
couples modes with $l'=l, l\pm 2$. The set of modes that constitutes the
eigenspace $K$ must follow selection rules (1), (2) and (3) and the
quasi-degenerate condition $|\omega_{nlm}-\omega_{\textrm{ref}}|<100~\mu$Hz.
The selection rule for reduced matrix elements due to large scale poloidal flow
to be non zero is
\begin{enumerate}
\setcounter{enumi}{3}
\item $l+l'+s$ must be even.
\end{enumerate}
In this case the set of modes in the eigenspace $K$ must follow selection rules
(1), (2) and (4) and the quasi-degenerate condition as before. Note that the
matrix elements for rotation are real and symmetric whereas those for poloidal
flow are imaginary and antisymmetric. The perturbation matrix
$[\mathcal{H}_{n'n,l'l}^{m'm}]$ is hermitian for both cases and will have real
eigenvalues.
\section{Results}

\subsection{Differential rotation}
We show in this section that QD treatment would give non-zero even coefficients
$a_{2q}$ even in absence of the centrifugal term and their effect on $a_{2q+1}$
is indeed small as argued by Lavely \& Ritzwoller (1992).
The determination of even splitting coefficients due to differential rotation is
important as this may be used to remove the effect of rotation from the
observed splitting coefficients to isolate the effect of magnetic field and
other large scale flows in the observed splitting coefficients.
However, if DPT is used, only the odd coefficients $a_{2q+1}$ are non-zero if we
neglect the centrifugal force  which is second order in $\Omega$.
Further, in this approximation only the North-South symmetric component of
rotation gives non-zero frequency shifts.
Lavely \& Ritzwoller (1992) applied the QDPT to differential rotation to find
that for intermediate $l$ modes ($l \sim 50$) the quasi degenerate coupling will
have little effect on the modal frequencies and can be ignored for all practical
purposes.

It is easy to verify from equation~(\ref{eq:genmat}) that for pure rotation
(${\bf v}=\Omega (r, \theta)r\sin\theta \hat{\phi}$) the symmetric matrix
elements are odd functions of $m$. Hence from equation~(\ref{eq:m}) and
(\ref{eq:qdp4}) one can infer that the perturbed frequency will have both odd
and even splitting coefficients.
It can be easily shown that there will be no additional effect in QDPT due to
component $w_1^0$ of rotation velocity as the off-diagonal elements in
the resulting matrix would vanish. Thus the leading effect of rotation
arises from $w_3^0$ term. We calculate this contribution by using
\begin{equation}
w_3^0(r)=\cases{34.9 \textrm{ m s$^{-1}$} & if $r\ge 0.7\Rs$\cr 0, &
otherwise.\cr}
\end{equation}
This value is chosen such that the DPT gives a value of $a_3$ close to the
observed value for most modes trapped in the convection zone.
The QDP treatment modifies the odd splitting coefficient $la_1$ by up to 0.4 nHz
(see Fig.~\ref{fig:rot1}a) and $l a_3$ by up to 0.2 nHz, which are a couple of
orders of magnitude smaller than the typical errors in these coefficients. 
In Figure~\ref{fig:rot1}b we plot the splitting coefficient $l a_2$ for all the
modes with frequency less than 4.5 mHz and $l\ge 10$, against $r_t^{nl}$, the
lower turning point of the mode.

The calculation has been done using an equation similar to
equation~(\ref{eq:qdp4}) but for all the modes in the neighborhood of radius
$100~\mu$Hz about the central mode. The mathematical details of the calculation
has been discussed in  \S3.2 (see equations~(\ref{eq:matrot}) \&
(\ref{eq:matrot2})). The maximum value of $l a_2 \sim 30$ nHz which may be
significant depending upon the magnitude of the large scale flow or magnetic
field perturbations.  Note that the effect of centrifugal force on $la_2$ also
happens to be of the same order (see Fig.~5 of Antia, Chitre \& Thompson 2000).
We have emphasized earlier that in order to spot the signatures of the large
scale flows and magnetic fields we need to remove the effect of rotation on
$a_2$.

It has been found from Doppler measurements that there exists a hemispheric
asymmetry in surface rotation having an angular dependence
$\partial_{\theta}Y_2^{0}$ (e.g., Hathaway et al.~1996).
This component has an amplitude of $w_2^0 = -7.8 \pm 0.3$ m s$^{-1}$, which
indicates that the southern hemisphere was rotating slightly faster than the
northern hemisphere during the period of the data. We have also calculated the
splitting coefficient $l a_2$ due to north-south asymmetry assuming it is
constant throughout the convection zone and they happens to be small with
$la_2\la 0.1$ nHz.

\subsection{Meridional circulation}
In this section we investigate theoretical frequency shifts due to meridional
circulation.  We start with $s=2, t=0$, which
is the dominant component in the observed meridional flow near the
surface.
In our analysis, we have included all f- and p-modes with frequency less
than 4.5 mHz and $l\ge10$.
We consider a meridional circulation with one cell in each hemisphere and
maximum horizontal velocity at the surface equal to 30 m s$^{-1}$ which is
consistent with the Doppler measurements and ring diagram analysis.
This gives $u_0 = 9$ m s$^{-1}$ in equation~(\ref{eq:profile}). From
equations~(\ref{eq:matele2}) \& (\ref{eq:matele3}) we see that the coupling
matrix for poloidal flows is Hermitian with zero diagonal elements. This means
that degenerate perturbation treatment cannot to be applied to calculate the
frequency shifts and it becomes essential to use the QDPT, unlike for rotation
where QDPT just provides a second order correction. In presence of zero diagonal
elements of the matrix $[\mathcal{H}_{ij}]$, 
equation~(\ref{eq:qdp4}) for the frequency shift for coupling between two modes
simplifies to,
\begin{equation}
\label{eq:del}
\delta\nu = \frac{\omega'_2 -\omega_2}{2\pi} \sim
\frac{{\mathcal{H}}_{12}^2}{4\pi\omega_2\Delta}\;.
\end{equation}
It may be noted that the sign of the frequency shift for the central multiplet
with $\omega_{\textrm{ref}}$ depends on the sign of $\Delta =
\omega_{\textrm{ref}}^2-\omega_1^2$, where $\omega_1$ is the nearest mode to the
central frequency. 
If there are many modes with frequency close to frequency of the central mode,
then we can expect these contributions to be added. In this case, if the
sign of frequency differences are not the same, the terms will partially
cancel each other. If for some pair of modes which satisfy the selection
rules, the frequency difference is very small, then their frequencies would
be shifted significantly.

In Figure~\ref{fig:sym} we show an example of the frequency shift $\delta
\nu(m)$ of the multiplet $(n,l)=(1,292)$, which was also considered by Roth \&
Stix (2008).
The Wigner $3j$ symbol with $t=0$ (implying $m'=m$) in
equation~(\ref{eq:genmat}) is an odd function of $m$ and so $\delta \nu(m)
\propto \mathcal{H}_{12}^2$ is symmetric about $m=0$ for all the multiplets.
Another important point to note from equation~(\ref{eq:del}) is that $\delta \nu
\propto u_0^2$. This has also been verified by numerical calculations.
The maximum shift we get for this multiplet from our calculations is $\sim 8.5$
nHz in contrast to Fig.~2 of Roth \& Stix (2008) where they obtain a maximum
shift of $0.1~\mu$Hz. 
On comparison with their equation~(15) we find a difference of a factor of 2 in
the first term in the expression of $H_s(r)$. Even after using their version of equation~(15) in our
calculations we were not able to reproduce the large shifts reported by them.
The reason for this discrepancy is not clear. One possibility is normalization
of velocity amplitude. The velocity decreases rapidly with height and if
the normalization is applied at higher level, then the entire velocity profile
will be scaled up.
In these calculations we have normalized $u_0$ by comparing it with the surface
velocity. We find that the peak value of $v_s^t(r)$ in Figure~\ref{fig:profile}
occurs a little below the surface.
For instance for a flow  with $s=2$ and a velocity of 30 m s$^{-1}$ near $r
=\Rs$, this peak in $v_s^t(r)$ happens to be 72 m s$^{-1}$ at $r=0.996\Rs$.
However, in reality such a rapid increase in $v_s$ hasn't been found from local
helioseismology which can probe up to a depth of 0.96$\Rs$. Hence, in our
opinion the theoretical frequency shift calculated using such a radial profile
of $v_s^t(r)$ is an upper limit to the actual shift. 
Ideally, we should normalize $u_0$ to get the maximum value of the horizontal
velocity to be
30 m s$^{-1}$. That will bring down the splitting coefficients by a factor
of 4 or more.
On the other hand,
if we normalize the velocity further up then for the same value of velocity the
calculated
splitting coefficients would go up. However, that is not realistic as in
that case the maximum velocity would be much larger than the observed value.
In Figure~\ref{fig:merd}a,b we plot the frequency shifts averaged over $m$ for
each multiplet where as 
Figure~\ref{fig:merd}c shows the splitting coefficient $l a_2$ 
calculated for all the multiplets.

We consider only even values of $s$ so that the meridional flow does not have
cross equatorial components. As we increase $s$, selection rules allow more and
more multiplets in a radius of $100~\mu$Hz to couple and there are many more
instances of `near degeneracy', so that the $\delta \nu \propto 1/\Delta$ in
equation~(\ref{eq:del}) for some multiplets becomes quite large. For example the
multiplets $(17, 56)$ and $(16, 64)$ which have a frequency difference of
$-0.06~\mu$Hz can be coupled by the meridional flow with $s=8$ to give $l a_2
\sim 480$ nHz.
In these calculations we have used $u_0=100$ m s$^{-1}$. In Figure~\ref{fig:sym}
we have plotted the frequency shifts for the pair of multiplets $(15, 34)$ and
$(14, 40)$ also considered by Roth, Howe \& Komm (2002), coupled by the same
flow. The result for all multiplets with frequency less than 4.5 mHz and
$l\ge10$ is shown in Figure~\ref{fig:merds8}a,b.
It is to be noted from Figure~\ref{fig:merds8}b that only some multiplets have
$a_2$ larger than the errors in $a_2$ for those modes in observational data. It
is only these multiplets which we may hope to detect in the observations. 
In this case we have taken the errors from the GONG data set centered at 19
November 2002. We present comparisons with GONG as well as MDI data in \S4.4.
Apart from the nearly degenerate modes the splitting coefficients of other
modes are larger than that found for $s=2$ with similar value of horizontal
velocity near the surface.
It may be noted that for large values of $s$ the maximum value of horizontal
velocity would be quite different from $v_s^t(r)$ because of the additional
factor arising from the gradient of spherical harmonics (cf., Eq.~\ref{eq:flow}).
In the case where we normalize $u_0$ such that $v_s^{0}(\Rs) = 30$ m s$^{-1}$, we also obtain an increase with $s$ in the splitting coefficients as reported by Roth \& Stix (2008).
This increase is not seen if the maximum value of the horizontal velocity near
the surface is normalized to 30 m s$^{-1}$.
Of course, observations near the solar surface do not show such large
magnitudes for these higher order components of meridional flow.
If realistic amplitudes are used then the effect will be negligible.

\subsection{Giant convection cells}
In this section we calculate the splittings due to poloidal flows with angular
dependences that depend on longitude, $\phi$.
We consider the flow given by spherical harmonics $Y_8^{4}(\theta, \phi)$ and
$Y_8^{8}(\theta, \phi)$ (also  called sectoral rolls or banana rolls).
This calculation is little more involved than that in \S4.2 as a non zero $t$ allows
coupling between different $m$ and $m'$ in equation~(\ref{eq:genmat}). Hence it
becomes important to take into account the effect of rotational splitting on
$\Delta$ before calculating the effect of these kind of poloidal flows. It is
easy to see from the properties of the Wigner $3j$ symbols that the $s=8,t=8$
flow couples the mode $(n,l,m)$ with $(n',l',m\pm 8)$. Hence the difference of
the square of frequencies $\Delta$ is no longer independent of $m$  unlike in
\S4.2.

Let us consider the perturbation in the frequency of the multiplet $(15,34)$.
The nearest multiplet to this happens to be $(14,40)$ with a frequency
difference of $4.12~\mu$Hz and the next one is $(16,28)$ with frequency
difference $19.01~\mu$Hz.
This is the same mode shown by Figure~3 of Roth, Howe \& Komm (2002).
We use the rotational splittings calculated from temporally averaged rotation
rate as inferred from the GONG data to calculate $\Delta$.
The result of our calculation including only the nearest multiplet $(14,40)$ has
been shown in Figure~\ref{fig:asym1}. The asymmetric frequency shift shown by
the solid line in Figure~\ref{fig:asym1}a denotes the change in the frequency of
the mode $(15, 34, m)$ due to coupling with the mode $(14, 40, m-8)$ whereas the
dashed line denotes the change due to coupling with $(14, 40, m+8)$. As shown in
equation~(\ref{eq:del}) a crucial component of the shift comes due to the
difference of the squared frequencies  $\Delta_{m,m'}$ of the modes coupling
according to the QDPT condition and the selection rules.
The corresponding $\Delta_{m,m-8}$ and $\Delta_{m,m+8}$ are plotted in
Figure~\ref{fig:asym1}b. The total shift is given by the sum of the solid and
dashed curves.
The couplings within a multiplet i.e.,  $(n,l,m) \rightleftarrows (n,l,m \pm 8)$
happen to be zero because of the anti-symmetry of the matrix elements. The
values of splitting coefficients $l a_1$ and $l a_2$ for the mode $(15, 34)$ are
calculated to be $-12.2$ nHz and $-5.7$ nHz and that for $(14, 40)$ are 11.1 nHz
and 5.3 nHz, respectively.

We repeat this calculation for all modes with frequency less than 4.5 mHz and
find that some of the multiplets have quite high values of $a_2$ (see
Fig.~\ref{fig:a1rt77}b). For example, the multiplet $(18, 61)$ which couples
with nearest multiplet $(17, 69)$ with a frequency difference of $-3.4~\mu$Hz.
In Figure~\ref{fig:asym2}a we have plotted $\delta\nu(m)$ versus $m$ for the
coupling $(18, 61, m) \rightleftarrows (17, 69, m-8)$ which seems to have a
discontinuity at $m=-22$. This is due to the fact that $\Delta_{m,m-8}$ for this
coupling also changes sign at $m=-22$.
Note that two modes $(18, 61, m)$ and $(17, 69, m-8)$ having a mean frequency
difference of $4.12~\mu$Hz become `nearly' degenerate at $m = -22$.
This jump presumably gives rise to high values of splitting coefficients. The
other coupling $(18, 61, m) \rightleftarrows (17, 69, m+8)$ is smooth and so is
the $\Delta_{m,m+8}$ as shown in Figure~\ref{fig:asym2}b by the dashed line. We
also calculate the total shift experienced by the multiplet $(18, 61)$ while
coupling with 3 nearest modes $(17, 69), (19, 55), (19, 53)$ and find no
significant difference with Figure~\ref{fig:asym2}a. We have also shown a
calculation for the asymmetric shifts for the multiplet $(14, 34)$ and $(14,
40)$ coupled due to a flow varying as $Y_8^4(\theta, \phi)$ in
Figure~\ref{fig:asym4}.

The asymmetric shift due to these couplings give rise to non-zero odd splitting
coefficients in addition to the even coefficients.
Since rotation inversions neglect this contribution, the result may not be
correct if the giant cells have significant amplitudes. The effect of giant
cells on rotation inversions can be estimated by performing rotation inversion
using odd splitting coefficients from these calculations. The resulting rotation
rate should be subtracted from the actual rotation inversions.
It is likely that if these multiplets are used for helioseismic inversion using
the rotation kernel, they will give rise to distortion in the inverted rotation
profile.
The effect on the odd splitting coefficients, $a_1$ and $a_3$ is shown in
Figure~\ref{fig:a1rt77}a,c in terms of observational errors in those multiplets.
We use the theoretically calculated $a_1$ and $a_3$ to perform a 1.5d
helioseismic rotation inversion
using the formulation due to Ritzwoller \& Lavely (1991). 
We use 1.5d inversion as we have calculated only 2 splitting coefficients, $a_1$
and
$a_3$. For this purpose, we use only those modes
which are present in a GONG data set and use the errors in those modes
for inversions using a Regularized Least Squares method using the same
smoothing as used for inverting real data (e.g., Antia et al.~1998).
The results are shown in Figure~\ref{fig:inv}, which shows the resulting
$w_1^0(r)$ and
$w_3^0(r)$ in terms of the estimated errors. Because of presence of modes
with large $a_1,a_3$ with $r_t\approx 0.7R_\odot$, there are some oscillations
in this region, but their amplitudes are less than the estimated errors.
Thus the effect on rotation inversion may not be significant even when
the giant cells have velocity of 100 m s$^{-1}$. With increasing velocity
the effect would be larger and it may be possible to detect this effect
as it will give an oscillatory signal in rotation inversion at these depths.
This has been pointed out by Roth, Howe \& Komm (2002). They find a much larger
effect as their
estimate of frequency shift is generally larger than what we find.
In real data the effect may not be as dramatic as shown by Figure~8 of Roth,
Howe \& Komm (2002)
since many of the modes with large shifts in $a_{2q+1}$ would yield large
residuals
in inversions,
and any reasonable inversion would eliminate such modes. In our calculation,
we found residuals going to about $4\sigma$ level and did not eliminate
any modes. Thus it is difficult to put any limits on magnitude of flows
using inversion results.

The even coefficient $a_2/\sigma_2$ is also shown in Figure~\ref{fig:a1rt77}b. 
If we find some multiplets with $0.6<r_t/\Rs <0.8$ having magnitudes larger than
the errors in observations as in  Figure~\ref{fig:a1rt77}b we should be able to
throw light on the flow velocities in giant cells. We shall look for such
features in \S4.4.

\subsection{Comparison with observed data}
We use 7 different data sets from the GONG observations, covering the late
descending part of cycle~22 to end of cycle~23, i.e., from June 1995 to November
2008. We choose only those modes for which $a_2$ is available in the data sets.
Since the different data sets can have different set of modes we repeat the
calculation for all the 7 data sets.
Most of the contribution to observed splitting coefficients appear to
come from near surface effects, while we are interested in modes with
turning point close to the base of the convection zone.
Thus we first separate the surface effects in the data by fitting a cubic spline
in terms of frequency to $lI_{nl}a_{2k}/I_{nl0}Q_{lk}$ versus frequency.
Here, $I_{nl}$ is the mode inertia of the mode (e.g., Christensen-Dalsgaard 2002),
while $I_{nl0}$ is the mode inertia
for the $l=0$ mode at the same frequency, while $Q_{lk}$ is a geometric factor
as defined by Antia et al.~(2001).
After subtraction of the surface effect from the observed splitting
coefficients, all the multiplets are sorted in the order of increasing $r_t$ and
then an error weighted 100 point average is applied. In
Figure~\ref{fig:a2rt77}a, b we show the result of the averaging for the GONG
108-day averaged data sets centered on 19 November 2002 and 18 September 2007
respectively.
To get an idea of the magnitude of the giant cell flows one 
can consider comparing the size of the valley and the hump in the region
$0.6<r_t/\Rs<0.8$ in the theoretical curve with that in the observation.
It is difficult to make out if the expected signature is present in the
observed data sets. In most cases no significant hump is seen in this region,
though in some cases like in Figure~\ref{fig:a2rt77}a there is some hint of
hump which is comparable to errorbars.
The main problem with this approach is that the running mean tends to average
modes with positive and negative values of $a_2$, thus reducing the
significance of the results. Thus it may be better to separate out these
modes to look for the signal.

We have said earlier that only way to put an upper limit on the flow velocity is
to look for signatures of nearly degenerate modes in the data.
In order to do that we divide the multiplets from the theoretical calculations
in two groups: one for which $l a_2 > 10$ nHz and the other with  $l a_2 < -10$
nHz.
Amongst 3700 theoretically evaluated multiplets we find only about 250
multiplets satisfying either criteria. Then we further search for these
multiplets in the GONG data sets and perform an averaging over the available
multiplets to get an estimate of $a_2$ from theory and error $\sigma_2$ in $a_2$
from the observations.
Only about 10--12 of the 250 multiplets are usually found to be present in the
GONG data sets.
We find that for all the data sets, $|l a_2| \sim 22$ nHz for theoretically
calculated coefficients, whereas the error in $l a_2$ (in observed set)
is $\sim 14$ nHz for the $Y_8^4(\theta,\phi)$ flow with $u_0=100$ m s$^{-1}$.
The corresponding values for the $Y_8^8(\theta,\phi)$ flow with a $u_0=50$ m
s$^{-1}$ are 28 nHz and 13 nHz respectively.
Since the observed data do not show these values (in both $>10$ nHz and $<-10$
nHz sets) we can only put an upper limit on velocities in such flows.
For this purpose we compare the $a_2$ values from theory with observational
errors to calculate the confidence level of the upper limit.
So we can say that up to a $1.5\sigma$ level we can rule out the
$Y_8^4(\theta,\phi)$ dependent flow with $u_0>100$ m s$^{-1}$ and the existence
of the $Y_8^8(\theta,\phi)$ flow with $u_0>50$ m s$^{-1}$ can be eliminated with
a confidence level of $2 \sigma$.
We have also performed some calculations with $Y_8^3(\theta,\phi)$ angular
dependence and can say with a $1.5\sigma$ confidence level that such flows also
cannot have $u_0>70$ m s$^{-1}$. Our results for multiplets in all the 7 GONG
data sets divided into two groups according to theoretical value of $la_2$  are
represented in Table~1. MDI data also give similar results.
By looking at detailed peak profiles of these abnormal modes in observed
data it may be possible to put more stringent limits on velocity of such
flows.

\section{Conclusions}
In this study we have calculated the effect of differential rotation, the
meridional circulation and the giant cell flows on p-mode frequencies using  
quasi-degenerate perturbation theory. For toroidal flows like differential
rotation, the quasi degenerate theory provides only a second order correction to
the eigenfrequencies, 
whereas for the poloidal flows it becomes necessary to use the quasi degenerate
treatment 
since the diagonal terms of the perturbation matrix vanishes. We agree with
Lavely \& Ritzwoller (1992) that the effect of rotation on the odd coefficients
is negligible and hence using the degenerate theory is sufficient.
However, the even splitting coefficients $a_{2q}$ are about one-half of the
errors in observations and they might not be negligible while calculating the
$a_{2q}$ for other effects like the magnetic field. 
This contribution is comparable to the effect of centrifugal term.

The frequency shift due to N--S antisymmetric component of rotation rate
vanishes in the
degenerate perturbation treatment and it is necessary to use QDPT.
We find that with realistic magnitude of velocities in the antisymmetric
component, the computed splitting coefficients are very small and this
effect may be neglected.

For a meridional 
circulation with maximum horizontal velocity at the surface of
$30$ m s$^{-1}$ and one cell per hemisphere we find that $l a_2 \sim 8.5$ nHz for
multiplets 
with turning points $r_t$ near the surface.
Because of our choice of velocity field, the horizontal velocity increases
rapidly with depth near the surface. This behavior is not seen in the Sun.
In that case it may be more meaningful to choose velocity profile with
a maximum value of 30 m s$^{-1}$, which occurs a little below the surface.
In that case the splitting coefficients would need to be reduced by a
factor of 4 or more, making them even smaller.
The mean frequency shift due to meridional flow is found to be up to 12 nHz,
which is much smaller than what is found by Roth \& Stix (2008).
The reason for this discrepancy is not clear. This could be due to differences
in where the velocity is normalized to specified value.

In any case, $la_2$ is very small for multiplets with $r_t/\Rs < 0.9$ coupled
with single celled meridional circulation. This value is less than roughly
one-seventh of the errors in the observed splitting coefficients. We also do not
find any change 
with increasing the number of cells in the radial direction or by changing the
depth of penetration of the meridional circulation while keeping the surface
amplitude constant. This makes it impossible to comment on the return flow
believed to be present at the base of the convection zone. When we increase the 
number of cells per hemisphere the selection rules allow more multiplets to 
couple with each other. In the process some of the nearly degenerate multiplets
combine to 
give rather large values of $l a_2$. Interestingly, most of these nearly
degenerate modes have turning points in the range $0.6\Rs$ to $0.8\Rs$. These
purely meridional flows give frequency shifts which are symmetric about $m = 0$
and only give non zero values for even splitting coefficients. 
The observed amplitude of these higher order components near the surface is
rather small and with such amplitudes the effect is expected to be small.
It appears that in general the magnitude of splitting coefficients due
to $u_0=9$ m s$^{-1}$ with $s=2$ is smaller than that with $u_0=100$ m s$^{-1}$
and $s=8$, even after excluding modes with nearly degenerate frequencies.
This could be due to a larger maximum horizontal velocity as well as larger
radial velocity (for $s=8$) or some geometric factors arising in the calculations.

In addition to this, we have also calculated the asymmetric frequency shifts
caused due to giant convection cells with angular flow  profiles given by
$Y_8^4(\theta, \phi)$ and $Y_8^8(\theta, \phi)$. Our results differ
significantly from earlier works of Roth, Howe \& Komm (2002) and Roth \& Stix
(2008) who find the effect of poloidal flows on p-mode frequencies to be an
order of magnitude larger than what we find.
The asymmetric shifts also contribute to the theoretically calculated odd
coefficients. We have performed a 1.5d rotation inversion on these coefficients
to detect any discernible feature in the inverted profile. However, since the
magnitude of the features in the inverted profile are smaller than the inversion
errors we conclude that giant cells with $u_0 \le 100$ m s$^{-1}$ do not have
much effect on the rotation inversion.

Finally we have analyzed some of the GONG data 
sets covering solar cycle~23 to look for possible evidence of giant cells.
We do not find any signal of these flows in selected modes with large
splitting coefficients and we can put upper limits on velocities of such
cellular flows.
From our analysis one can
say that the existence of giant convection cells of $Y_8^4(\theta, \phi)$
angular pattern with maximum vertical velocities of 100 m s$^{-1}$  can be 
ruled out with a confidence level of $1.5\sigma$ whereas cells with an angular
dependence $Y_8^8(\theta, \phi)$ cannot have vertical velocities $> 50$ m
s$^{-1}$ with a confidence level of $2\sigma$. It is important to remember here
that the GONG data sets are averaged over 108 days, while MDI data sets over 72
days.
This may be somewhat larger than the lifetimes of giant cells and hence
the signal may be averaged out. We can look at shorter data sets, but in
that case the errors would be larger. Further, the convective cells are
not expected to have smooth velocity profiles of the form used by us
over the entire solar surface. That would also reduce the contribution
to splitting coefficients. 

Finally, we would like to point out that while the
degenerate perturbation theory can be easily adopted for inversion because
of linearity of effect and the form of perturbation which can be represented
by appropriate kernels. In contrast, the effect of quasi-degenerate
perturbation theory is nonlinear in velocity and further is mainly governed
by frequency differences between nearly degenerate modes, rather than the
matrix elements which depend on velocity profile. Hence it is difficult
to use these shifts in inversions. For example, the main effect of
flow with $s=8$ is felt in modes with turning points close to $0.7\Rs$,
which happens to be close to the lower limit of region in which our flows
were localized. But we have verified that this is just a coincidence by
performing calculations with different lower boundary. The same set of
modes is affected irrespective of lower boundary, because it is these
modes that have small frequency differences. This is only determined by
the value of $s$ which determines the range of $l$ values that are
coupled. Thus various types of perturbations with the same coupling will
give large splittings in these modes and it is difficult to identify
the actual source of perturbation.

\acknowledgements

We thank the referee for useful comments and suggestions which have helped
in improving the manuscript. This work  utilizes data obtained by the Global Oscillation
Network Group (GONG) project, managed by the National Solar Observatory,
which is
operated by AURA, Inc. under a cooperative agreement with the
National Science Foundation. The data were acquired by instruments
operated by the Big Bear Solar Observatory, High Altitude Observatory,
Learmonth Solar Observatory, Udaipur Solar Observatory, Instituto de
Astrofisico de Canarias, and Cerro Tololo Inter-American Observatory.
This work also utilizes data from the Solar Oscillations
Investigation/ Michelson Doppler Imager (SOI/MDI) on the Solar
and Heliospheric Observatory (SOHO).  SOHO is a project of
international cooperation between ESA and NASA.
MDI is supported by NASA grants NAG5-8878 and NAG5-10483
to Stanford University.

\clearpage
\begin{figure}
%fig1
\label{fig:profile}
\includegraphics[width=0.9\textwidth]{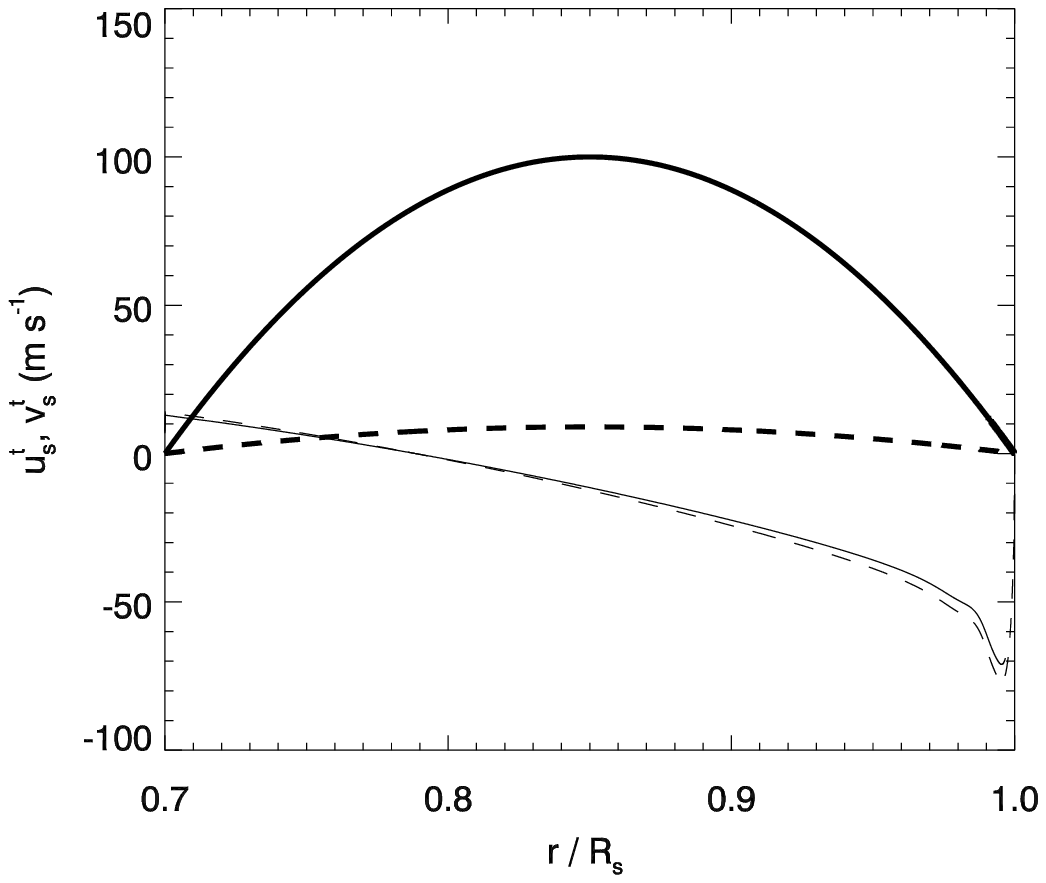}
\caption{Radial profiles of vertical velocity $u_8^0(r)$ (thick solid line) and
horizontal velocity $v_8^0(r)$ (thin solid line) with $u_0=100$ m s$^{-1}$ and
$u_2^0(r)$ (thick dashed line) and horizontal velocity $v_2^0(r)$ (thin dashed
line) with $u_0=9$ m s$^{-1}$. \label{fig:profile}}
\end{figure}
\clearpage
\begin{figure}
%fig2
\label{fig:rot1}
\includegraphics[width=0.9\textwidth]{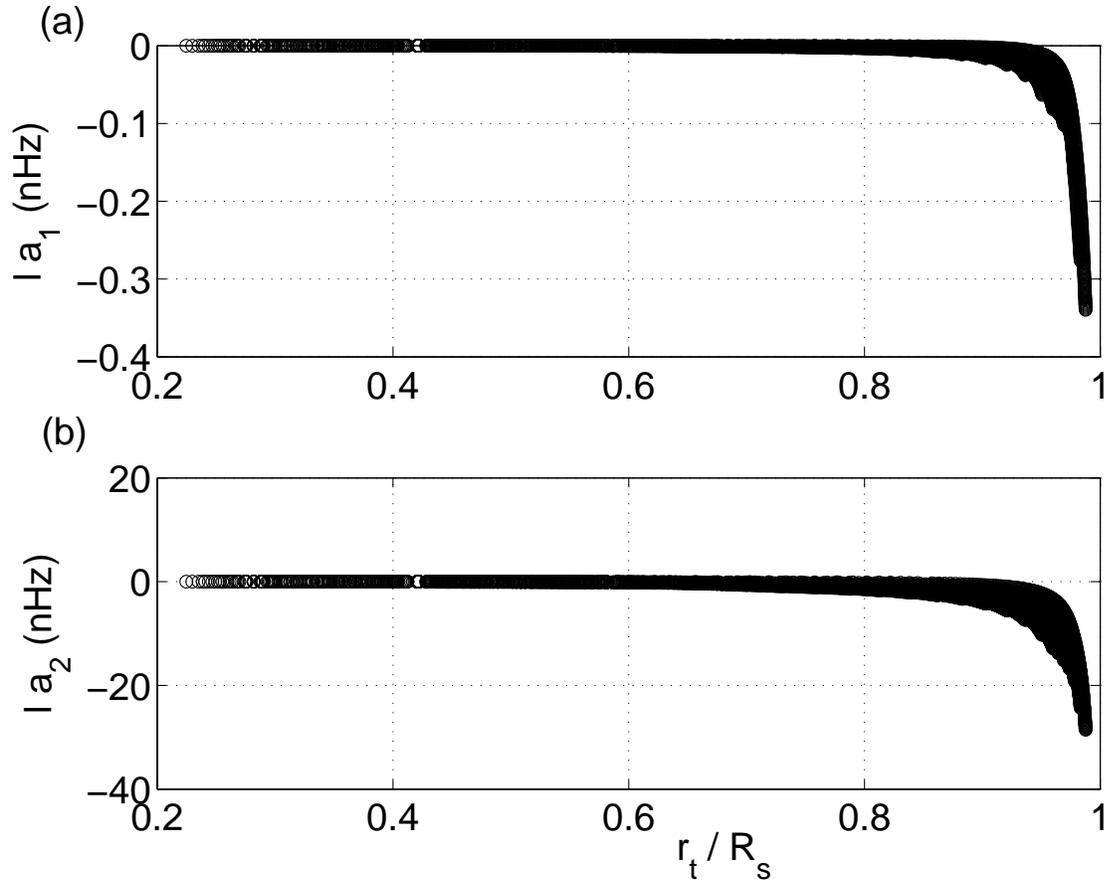}
\caption{(a) Odd splitting coefficients $l a_1$ as a function of $r_t$ for the
$\partial_{\theta}Y_3^0$ component of differential rotation (refer to
equation~(\ref{eq:flow})). (b) Same as (a) but for even splitting coefficient $l
a_2$. \label{fig:rot1}}
\end{figure}
\clearpage
\begin{figure}
%fig3
\label{fig:sym}
\includegraphics[width=0.9\textwidth]{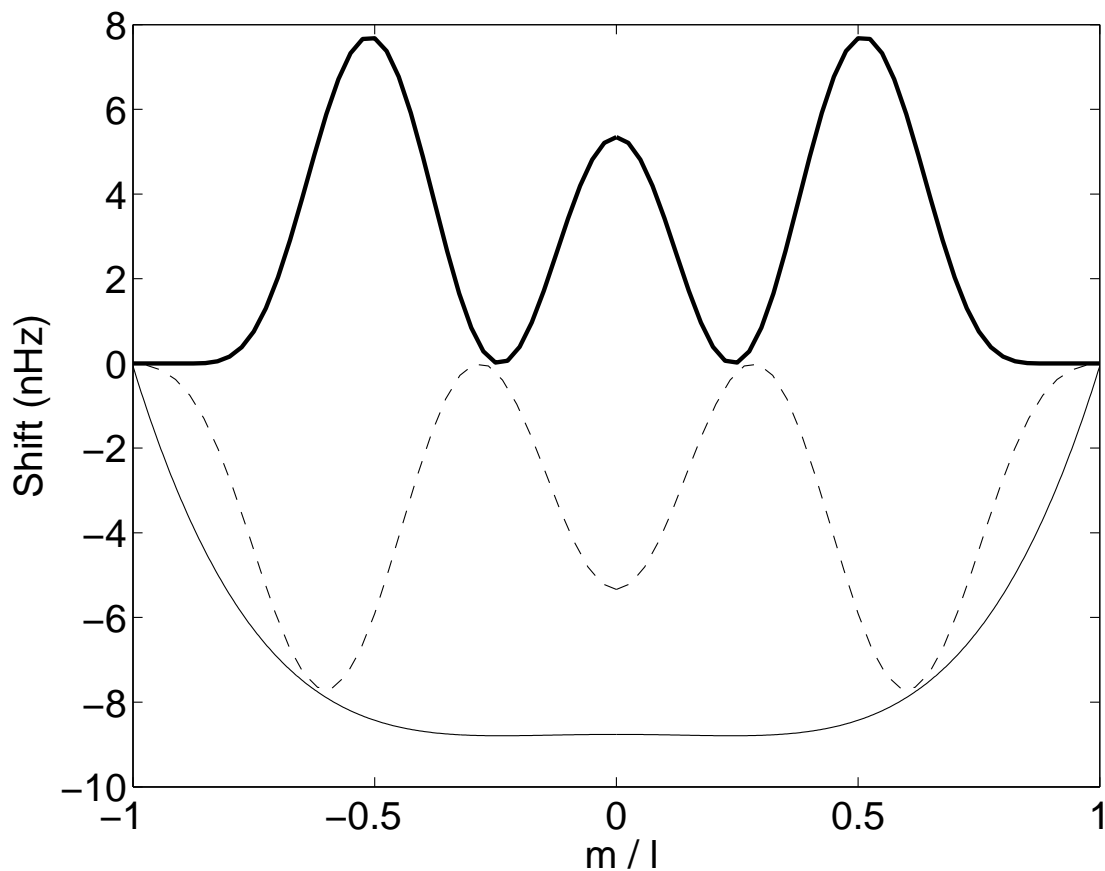}
\caption{Frequency shifts as a function of azimuthal order $m/l$. The shift in
the multiplet $(n,l) = (1,292)$ due to coupling by $Y_2^0$ flow with $u_0=9$ m
s$^{-1}$ (solid line);  $(n,l)=(15,34)$ due to $Y_8^0$ flow and $u_0=100$ m
s$^{-1}$ (dashed line); shift for the multiplet $(n,l)=(14,40)$ for the same
flow and velocity (thick solid line). \label{fig:sym}}
\end{figure}
\clearpage
\begin{figure}
%fig4
\label{fig:merd}
\includegraphics[width=0.9\textwidth]{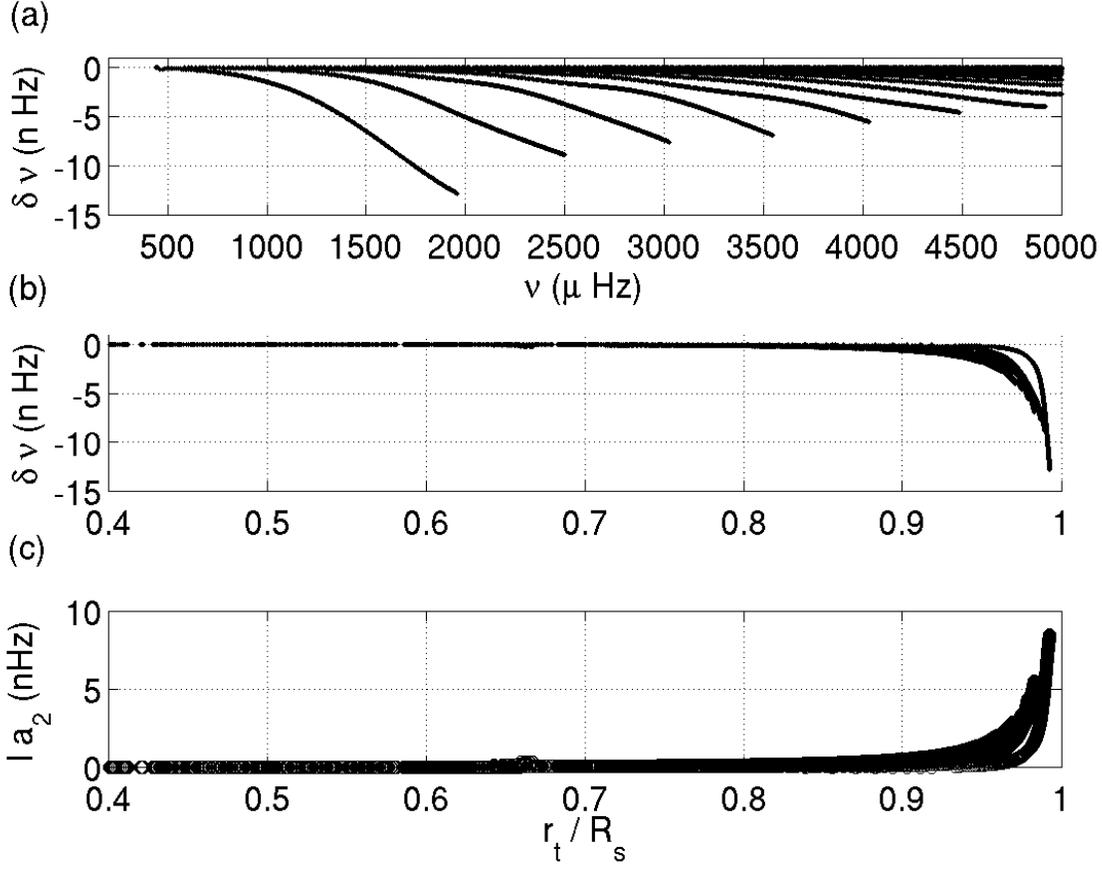}
\caption[]{\label{fig:merd} Mean frequency shifts $\overline{\delta \nu}(m)$ as
a function of (a) frequency $\nu$ and (b) turning point $r_t/\Rs$ due to
coupling by meridional flow with $s=2, t=0$. (c) Splitting coefficient $l a_2$
as a function of lower turning point radius $r_t$.}
\end{figure}
\clearpage
\begin{figure}
%fig5
\label{fig:merds8}
\includegraphics[width=0.9\textwidth]{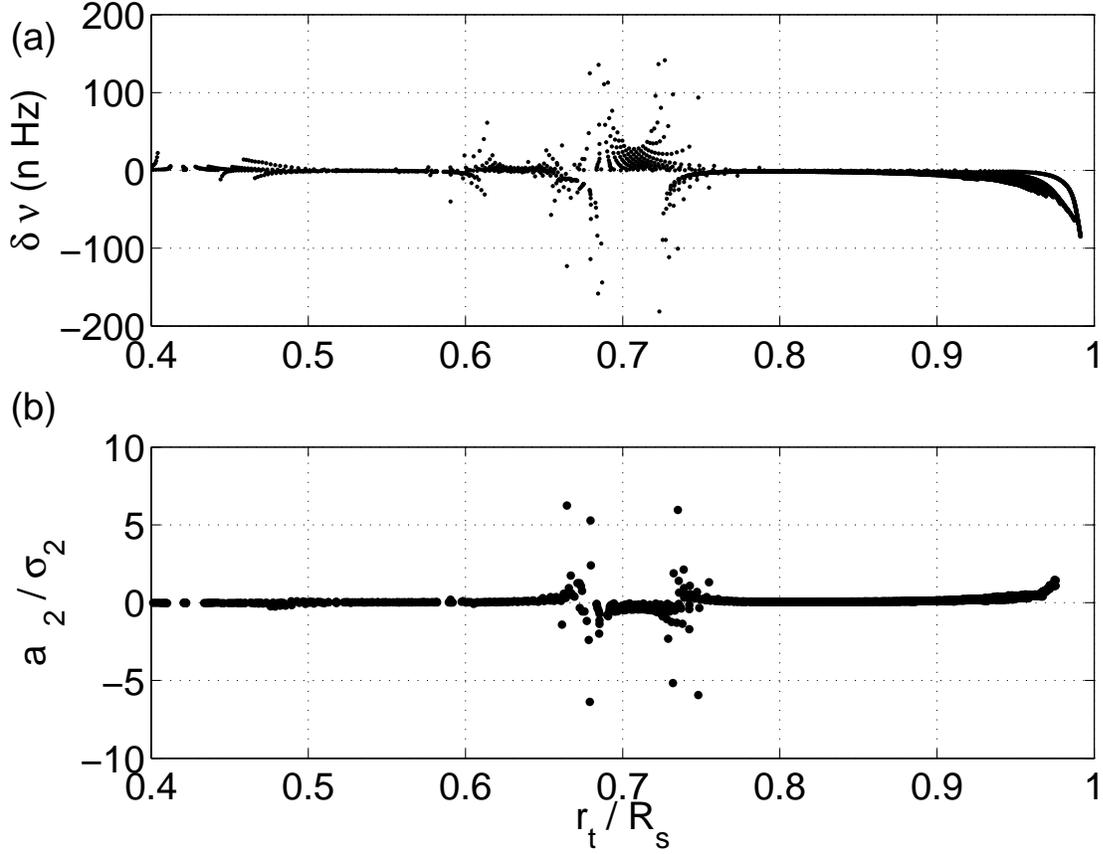}
\caption[]{\label{fig:merds8} (a) Mean frequency shift $\overline{\delta \nu}$
as a function of turning point $r_t / \Rs$ due to coupling by large scale flow
with $s=8, t=0$. (b) Splitting coefficient $a_2$ as a function of turning point
radius $r_t$. The points denote the theoretical values of $a_2/\sigma_2^{nl}$
for modes $(n,l)$. The errors $\sigma_2^{nl}$ have been chosen from the GONG
data set centered around 25 December 2002.}
\end{figure}
\clearpage
\begin{figure}
%fig6
\label{fig:asym1}
\includegraphics[width=0.9\textwidth]{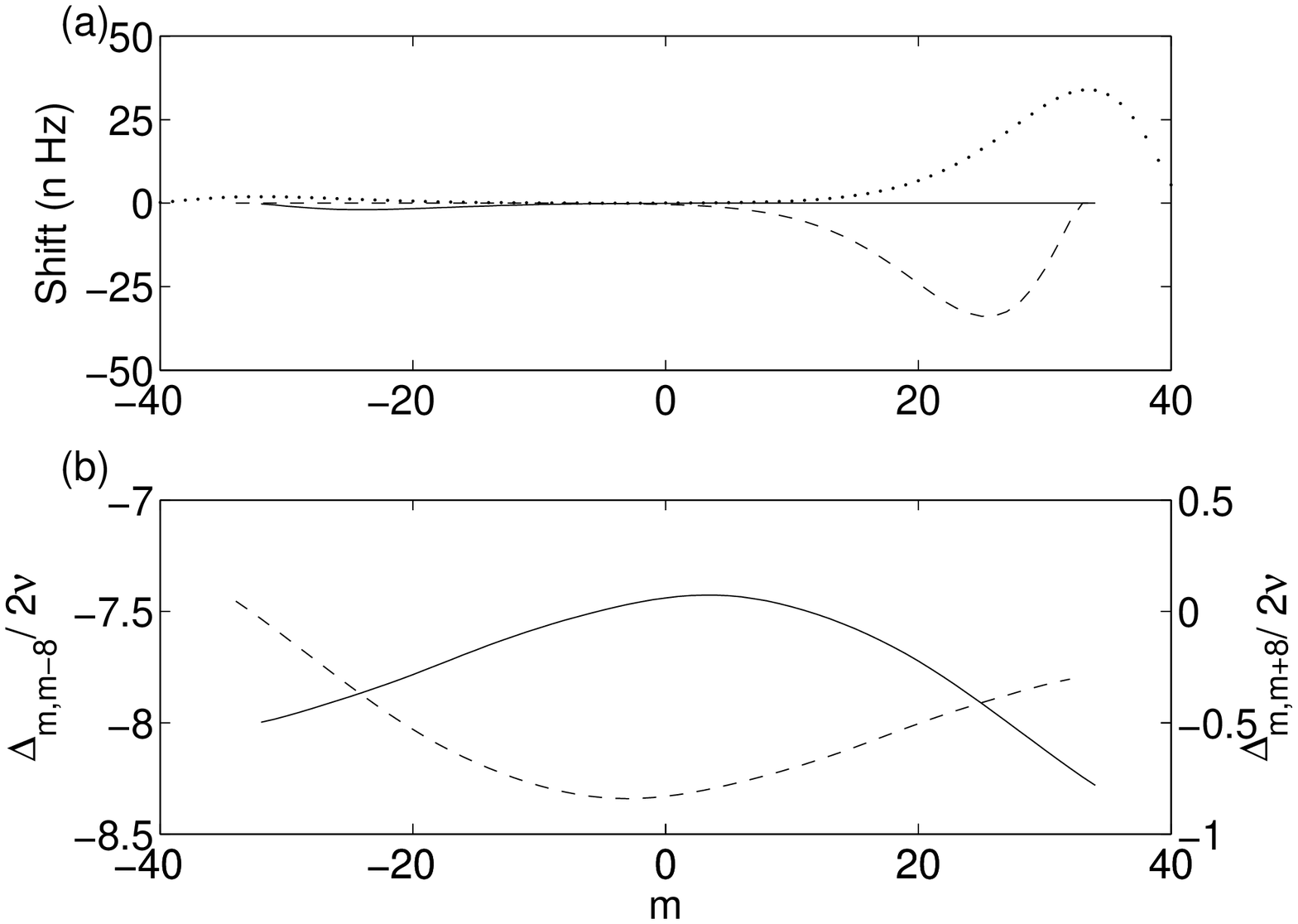}
\caption[]{\label{fig:asym1} (a) The {\em solid line} gives the frequency shift
because of interaction between modes $(15, 34, m)$ and $(14, 40, m-8)$ due to
the flow with an angular dependence $Y_8^8(\theta,\phi)$. The {\em dashed line}
is the frequency shift due to interaction between modes $(15, 34, m)$ and $(14,
40, m+8)$.  The total shift in frequency for the mode $(15, 34)$ is the sum of
the solid and the dashed lines. The frequency shift for the mode $(14, 40)$ is
given by the {\em dotted line}.
(b) $\Delta_{m,m-8}/2\nu$ ({\em solid line}) for the coupling
$(15,34,m)\rightleftarrows(14,40,m-8)$; and $\Delta_{m,m+8}/2\nu$ ({\em dashed
line}) for the coupling $(15,34,m)\rightleftarrows(14,40,m+8)$. These values are
in $\mu$Hz.  }
\end{figure}
\clearpage
\begin{figure}
%fig7
\label{fig:asym2}
\includegraphics[width=0.9\textwidth]{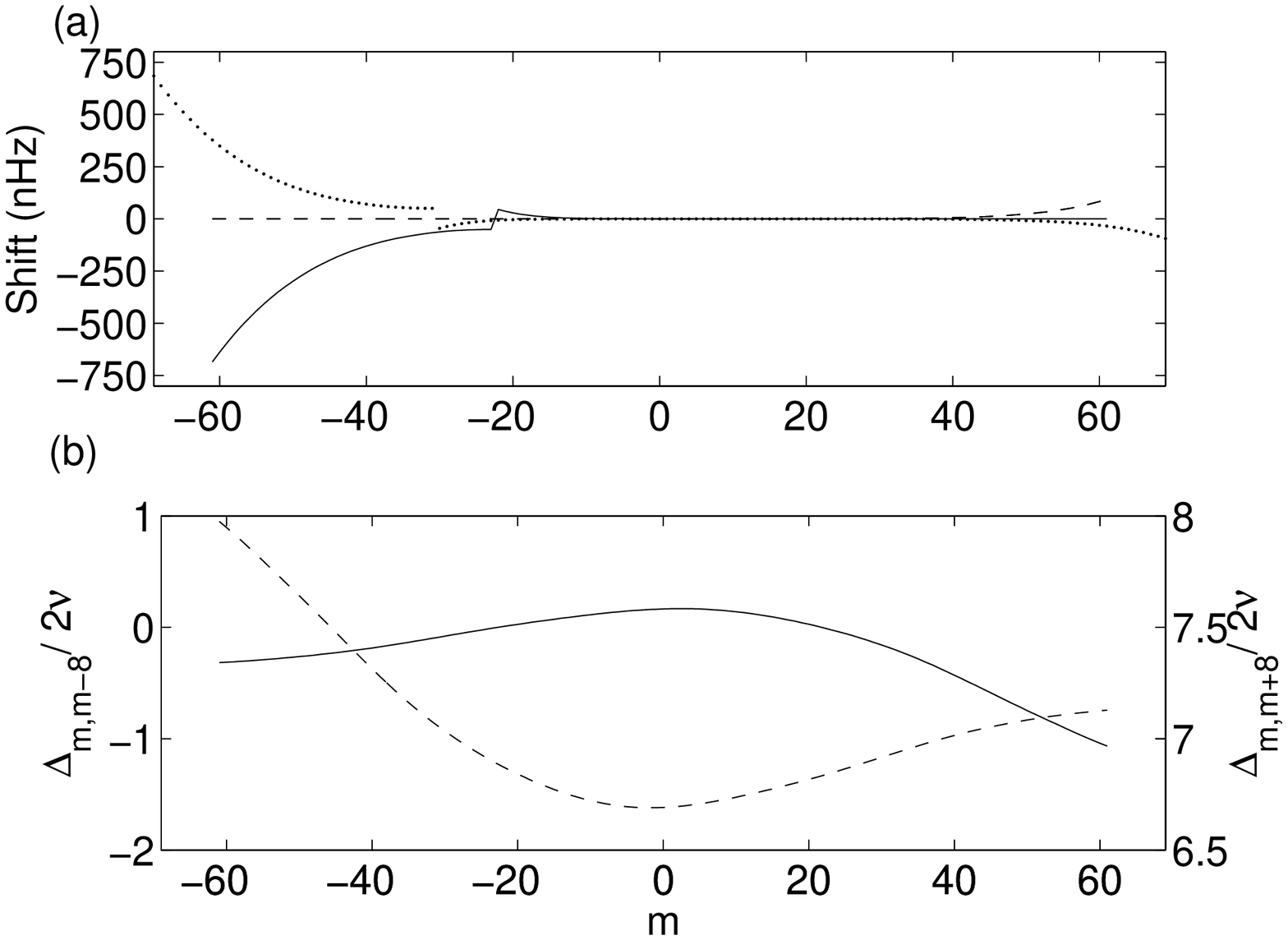}
\caption[]{\label{fig:asym2} (a) The {\em solid line} gives the frequency shift
because of interaction between modes $(18, 61, m)$ and $(17, 69, m-8)$ due to
the flow with an angular dependence $Y_8^8(\theta,\phi)$. The {\em dashed line}
is the frequency shift due to interaction between modes $(18, 61, m)$ and $(17,
69, m+8)$.  The total shift in frequency for the mode $(18, 61)$ is the sum of
the solid and the dashed lines. The frequency shift for the mode $(17, 69)$ is
given by the {\em dotted line}.
(b) $\Delta_{m,m-8}/2\nu$ ({\em solid line}) for the coupling
$(18,61,m)\rightleftarrows(17,69,m-8)$; and $\Delta_{m,m+8}/2\nu$ ({\em dashed
line}) for the coupling $(18,61,m)\rightleftarrows(17,69,m+8)$. These values are
in $\mu$Hz.}
\end{figure}
\clearpage
\begin{figure}
%fig8
\label{fig:asym4}
\includegraphics[width=.9\textwidth]{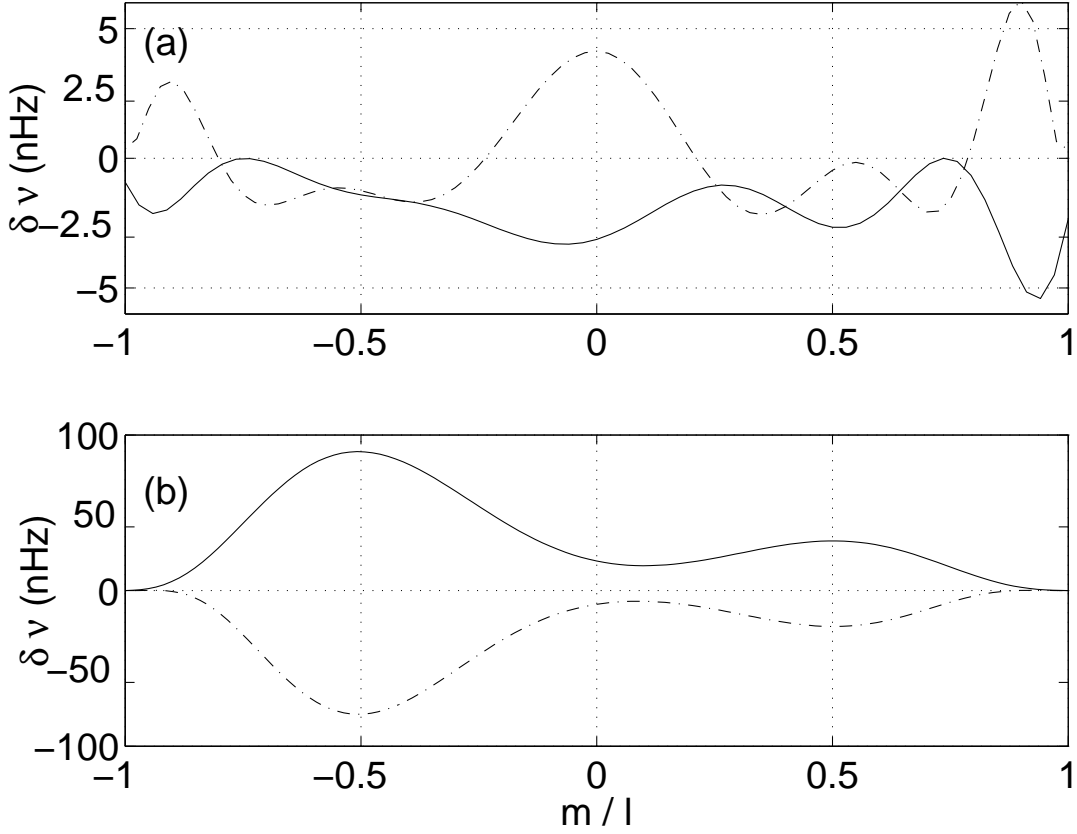}
\caption[]{\label{fig:asym4} Asymmetric shift in frequency of the multiplet (a)
$(15,34)$ (solid line) upon coupling with neighboring multiplets $(14,40)$ and
$(16,28)$ in presence of large scale flow $s=8,t=4$. The dashed-dotted line is the
shift for the mode $(14, 40)$ due to coupling with $(15, 34)$ and $(13, 48)$. 
(b) Same as (a) but for $(18,61)$ (solid line) upon coupling with neighboring
multiplets $(17,69)$ and $(19,55)$ in presence of large scale flow $s=8,t=4$.
The dashed-dotted line is the corresponding shift in $(17, 69)$ due to coupling with 2
nearest neighbors.}
\end{figure}
\clearpage
\begin{figure}
%fig9
\includegraphics[width=0.9\textwidth]{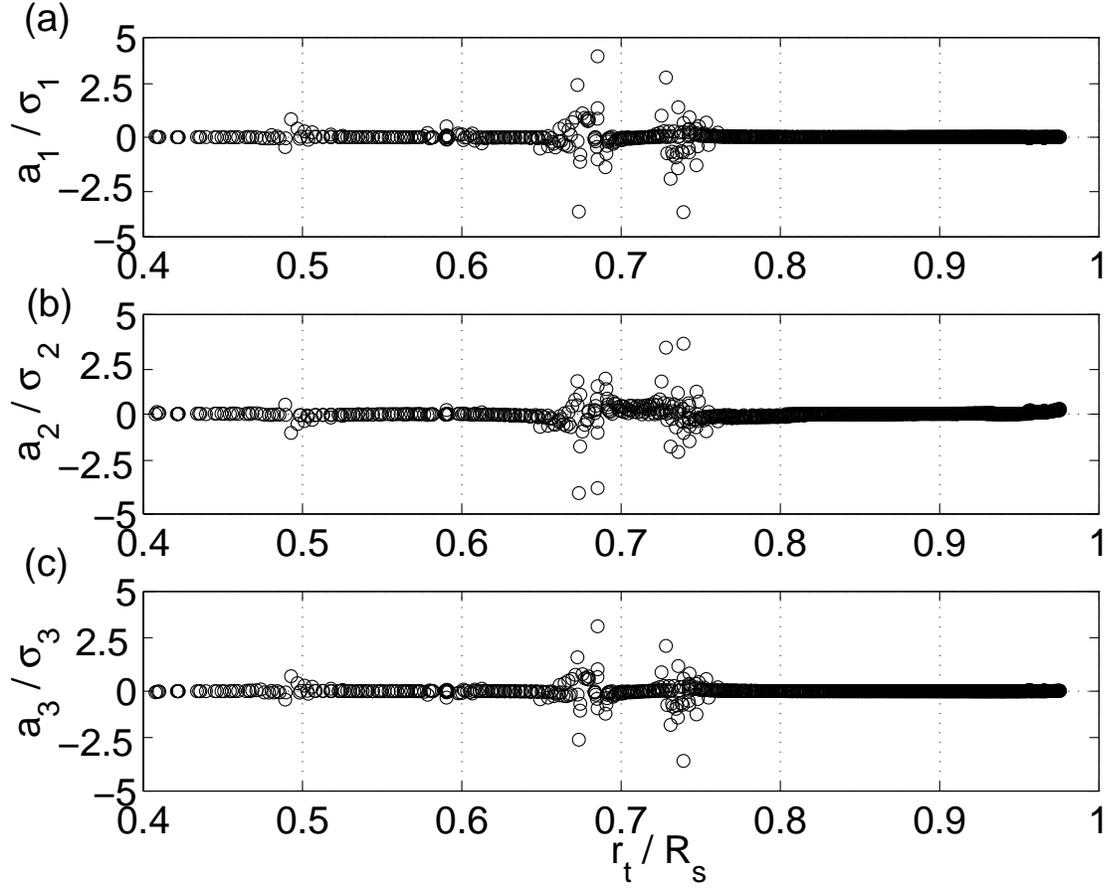}
\caption[]{\label{fig:a1rt77} (a) $a_1/\sigma_1$ and (b) $a_2/\sigma_2$ and (c)
$a_3/\sigma_3$ as a function of lower turning point radius $r_t$ for the
$Y_8^8(\theta, \phi)$ kind of flow with $u_0 =100$ m s$^{-1}$. The
$\sigma_{1,2,3}$ are the errors in the corresponding observational splitting
coefficients from the GONG data set centered about November 2002 (see
Table.~1).}
\label{fig:a1rt77}
\end{figure}
\clearpage
\begin{figure}
%fig10
\label{fig:inv}
\includegraphics[width=0.9\textwidth]{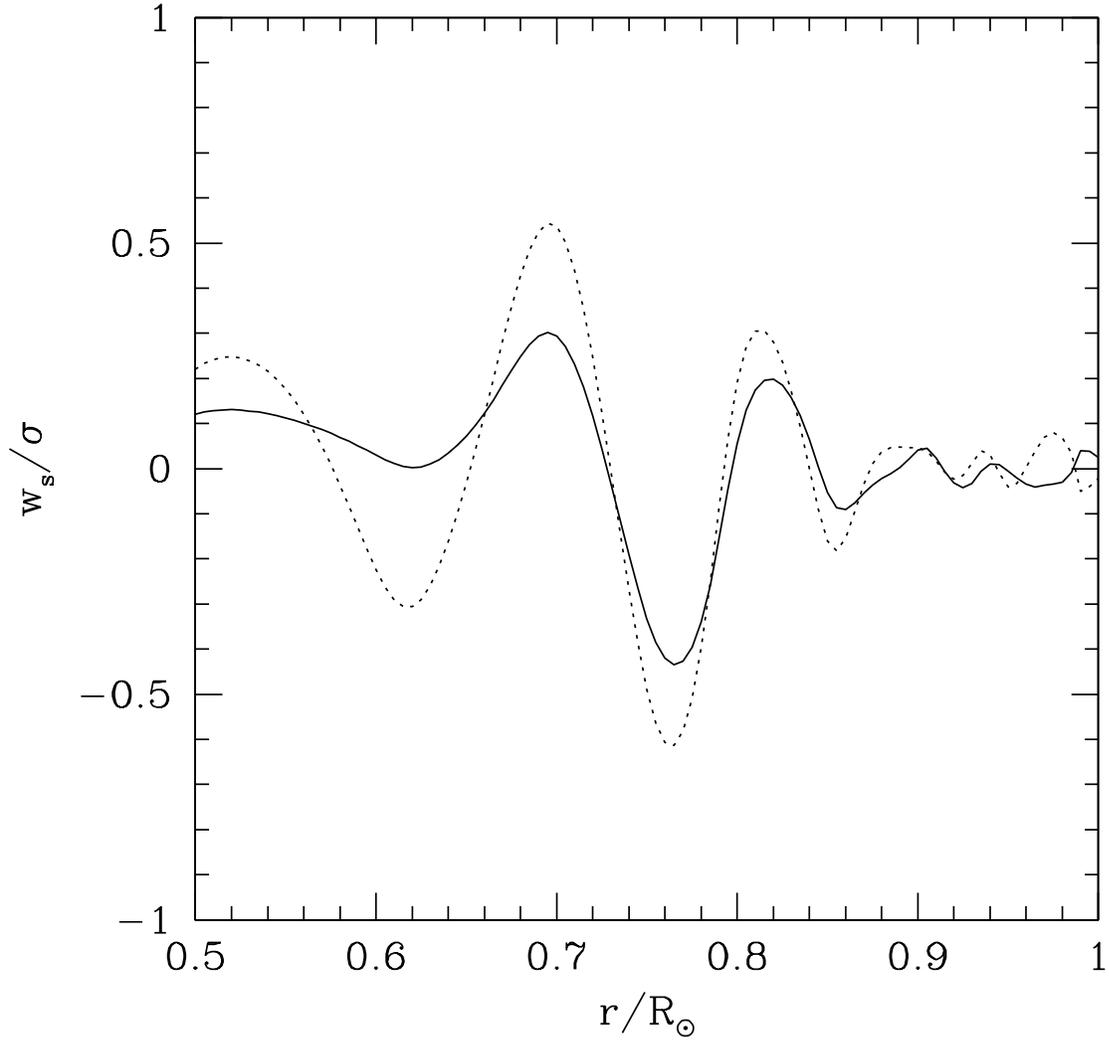}
\caption[]{\label{fig:inv} $w_1^0$ (solid line) and $w_3^0$ (dotted line)
in terms of the respective errors are shown as a function of radial distance
obtained from the 1.5d inversion of coefficients $a_1$ and $a_3$ shown in
Figure~9.}
\end{figure}
\clearpage
\begin{figure}
%fig11
\label{fig:a2rt77}
\includegraphics[width=0.9\textwidth]{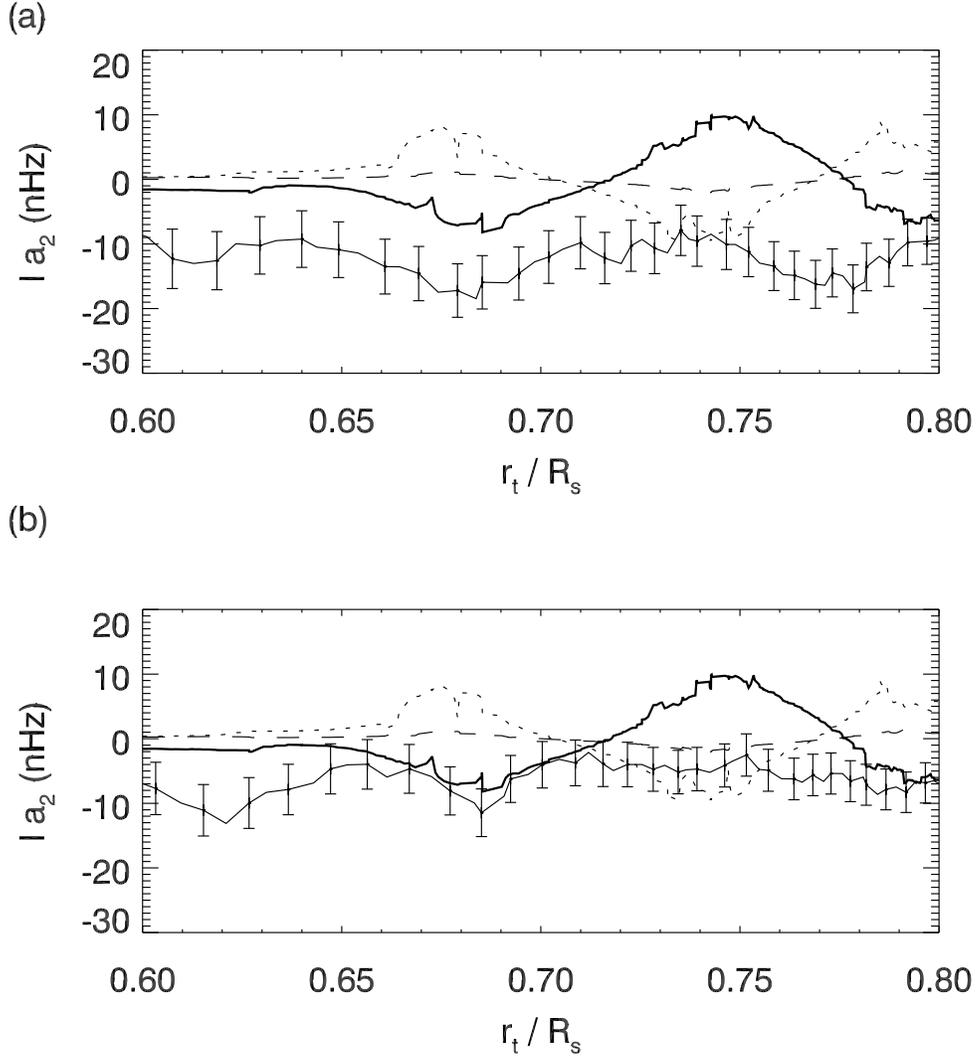}
\caption[]{\label{fig:a2rt77} $l a_2$ as a function of turning point radius
$r_t$ after averaging over 100 nearest points in $r_t$ for GONG data set
centered around (a) November 2002 and (b) 18 September 2007. The dotted line
correspond to the $Y_8^0(\theta, \phi)$ flow; dashed to $Y_8^4(\theta, \phi)$;
thick solid to $Y_8^8(\theta, \phi)$. The solid line with errorbars show the observed data points. To avoid congestion only a few representative errorbars are shown.}
\end{figure}
\clearpage
\begin{table}
\caption{Comparison of observed $l a_2$ with theory for flows with an angular
dependence $Y_8^t(\theta, \phi)$. The numbers in the GONG data set have the
following correspondence with the date about which the 108-day GONG month is
centered: (1) 29 June 1995, (2) 18 June 1997, (3) 18 September 2000, (4) 19
November 2002, (5) 23 July 2004, (6) 18 September 2007 and (7) 2 July 2008. For
$t=4$, $u_0 = 100$ m s$^{-1}$ whereas for $t=8$, $u_0 = 50$ m s$^{-1}$. Note
that for each $t$ the group I of multiplets correspond to $l a_2^\mathrm{th} <
-10$ nHz whereas the group II is for $l a_2^\mathrm{th} >10$ nHz. CL is the
confidence
level on the upper limit of $u_0$.}

\begin{tabular}{cc|cccc|cccc}
\hline 
 &  &  \multicolumn{4}{c}{I. $l a_2^\mathrm{th}<-10$ nHz} &
\multicolumn{4}{c}{II. $la_2^\mathrm{th}>10$ nHz} \\
\cline{3-10}
Data & $t$ & $l a_{2}^\mathrm{obs}$ & $l \sigma_{2}$ & $l a_{2}^\mathrm{th}$ &
CL & $l a_{2}^\mathrm{obs}$ & $l \sigma_{2}$ & $l a_{2}^\mathrm{th}$ & CL\\
set& & & & & ($\sigma$)& & & & ($\sigma$)\\
\hline
1   & 4 &17.7 & 16.9& -22.6 &1.3 & -3.7&14.8 &21.9 & 1.5\\
  & 8 & -11.3& 14.5&-29.0 & 1.9  & 3.3& 13.1&24.2 & 1.8\\
\hline
2   & 4  &-8.7 &14.9 &-22.2 & 1.5 &-12.4 &14.0 &21.3 & 1.5\\
  & 8 &-23.4 &13.2 &-28.1 & 2.1 &4.7 &11.7 &24.8 & 2.1\\
\hline
3   & 4 & -24.5&14.6 &-21.7 &1.5 & -21.8&14.3 &22.6 &1.6\\
   & 8 & -30.2&13.8 &-29.7 &2.1 & -17.7&12.3 &24.2 &2.0\\
\hline
4   & 4 & -1.3&14.5 &-22.3 &1.5 & -12.0&14.3 &21.8 &1.5\\ 
   & 8 & -30.2&13.8 &-29.7 &2.1 & -25.2&12.0 &23.9 &1.9\\
\hline
5   & 4 &-28.3 &14.2 &-22.0 &1.5& -5.8 &13.9 &22.0 &1.5\\
   & 8 & 5.3& 13.1&-27.3 &2.0 & -4.4& 11.8&24.1 &2.0\\
\hline
6   & 4 & -0.4&12.3 &-22.3 &1.8 & -22.1&12.8 &22.2 &1.7\\
   & 8 & -0.6&10.7 &-28.1 &2.6  & -4.5&10.4 &26.0 &2.5\\
\hline
7   & 4 & -13.2&14.2 &-22.2 &1.6  & -27.0&13.5 &21.8 &1.6\\
   & 8 & -27.4&12.7 &-28.9 &2.2 & -0.6&11.4 &24.6 &2.2\\
\hline
\end{tabular}
\end{table}
\label{lastpage}
\end{document}